%% file: paper.tex
\documentclass[conference]{IEEEtran}
\IEEEoverridecommandlockouts
\usepackage{cite}
\usepackage{amsmath,amssymb,amsfonts}
\usepackage{algorithm}
\usepackage{algorithmic}
\usepackage{graphicx}
\usepackage{subcaption}
\usepackage{romannum}
\usepackage{multirow}
\usepackage{textcomp}
\usepackage[dvipsnames,table,xcdraw]{xcolor}
\usepackage{lipsum}
\usepackage{listings}
\usepackage{todonotes}
\usepackage{hyperref}

\def\BibTeX{{\rm B\kern-.05em{\sc i\kern-.025em b}\kern-.08em
    T\kern-.1667em\lower.7ex\hbox{E}\kern-.125emX}}

\definecolor{blue-main}{rgb}{0,0,1}
\definecolor{dkgreen}{rgb}{0,0.6,0}
\definecolor{gray}{rgb}{0.5,0.5,0.5}
\definecolor{mauve}{rgb}{0.58,0,0.82}
    
\begin{document}

\title{Implementing Software Resiliency in HPX for Extreme Scale Computing\\\Large{Extended Abstract, SNL report number: SAND2020-3975 R}}


\author{
\IEEEauthorblockN{Nikunj Gupta\IEEEauthorrefmark{1}\IEEEauthorrefmark{4}, Jackson R. Mayo\IEEEauthorrefmark{2}, Adrian S. Lemoine\IEEEauthorrefmark{3}\IEEEauthorrefmark{4}, Hartmut Kaiser\IEEEauthorrefmark{3}\IEEEauthorrefmark{4}}
\IEEEauthorblockA{\IEEEauthorrefmark{1}\textit{Computer Science and Engineering} \textit{IIT Roorkee}, Roorkee, India, Email: gnikunj@cct.lsu.edu}
\IEEEauthorblockA{\IEEEauthorrefmark{2}\textit{Sandia National Laboratories}, Livermore, California, USA, Email: jmayo@sandia.gov}
\IEEEauthorblockA{\IEEEauthorrefmark{3}\textit{Center for Computation Technology}, \textit{Louisiana State University}, Baton Rouge, USA, Email: [aserio,hkaiser]@cct.lsu.edu}
\IEEEauthorblockA{\IEEEauthorrefmark{4}The STE$||$AR Group, http://stellar-group.org}
}


\definecolor{codegreen}{rgb}{0,0.5,0}
\definecolor{codegray}{rgb}{0.5,0.5,0.5}
\definecolor{codered}{rgb}{0.75, 0.3 ,0.1}
 
\lstdefinestyle{mystyle}{
    commentstyle=\color{codegreen},
    keywordstyle=\color{codered},
    numberstyle=\tiny\color{codegray},
    basicstyle=\ttfamily\footnotesize,
    breakatwhitespace=false,         
    breaklines=true,                 
    captionpos=b,                    
    keepspaces=true,                 
    numbers=left,                    
    numbersep=5pt,                  
    showspaces=false,                
    showstringspaces=false,
    otherkeywords ={hpxr::async_replay, hpxr::async_replay_validate, hpxr::dataflow_replay, hpxr::dataflow_replay, hpxr::dataflow_replicate, hpxr::dataflow_replicate_validate, hpxr::dataflow_replicate_vote, hpxr::dataflow_replicate_vote_validate, hpxr::async_replicate, hpxr::async_replicate_validate, hpxr::async_replicate_vote, hpxr::async_replicate_vote_validate},
    showtabs=false,                  
    tabsize=2
}

\lstset{style=mystyle}

\maketitle



\section{Introduction}
The DOE Office of Science Exascale Computing Project (ECP)~\cite{b1} outlines the next milestones in the supercomputing domain. The target computing systems under the project will deliver 10x performance while keeping the power budget under 30
megawatts. With such large machines, the need to make applications resilient has become paramount. The benefits of adding resiliency to mission critical and scientific applications, includes the reduced cost of restarting the failed simulation both in terms of time and power.

Most of the current implementation of resiliency at the software level makes use of a Coordinated Checkpoint and Restart (C/R)~\cite{b2, b3, b4, b5, b6, b7}. This technique of resiliency generates a consistent global snapshot, also called a checkpoint. Generating snapshots involves global communication and coordination and is achieved by synchronizing all running processes. The generated checkpoint is then stored in some form of persistent storage. On failure detection, the runtime initiates a global rollback to the most recent previously saved checkpoint. This involves aborting all running processes, rolling them back to the previous state and restarting them.

In its current form, the Coordinated C/R is excessively expensive on extreme-scale systems. This is due to the high overhead costs of global rollback followed by global restart. Adding to these overheads are the significant overheads of global I/O access. In many cases, millions of processes have to respond to a local process failure which leads to heavy loss of useful CPU computation cycles and leads to a significant performance penalty. This was observed when node level resiliency was implemented in a production application running on Titan system at Oak Ridge National Laboratory~\cite{b8}. The overheads of resiliency had a significant impact on performance as the overheads of C/R were 20-30\% of the total execution time.

Emerging resilience techniques, such as Uncoordinated C/R~\cite{b10} and Local Failure Local Recovery (LFLR)~\cite{b9} attempts to mitigate some of the overheads of coordinated C/R by eliminating the requirement of aborting all running processes and restarting. However, these techniques are based on assumptions exclusive to Single Program Multiple Data (SPMD) model i.e. the same program execution across all running processes. Asynchronous Many-Task (AMT) execution models provide similar resilience techniques without any of these assumptions.

In this paper, we explore the implementation of resiliency techniques in Asynchronous Many-Task (AMT) Runtime Systems. 
We have chosen to use HPX as a model AMT as it exposes a standards conforming API which is easy to understand and adopt. AMTs replace the bulk-synchronous MPI model with fine-grained tasks and explicit task dependencies. They rely on a runtime system to schedule the tasks and manage their synchronization. In an AMT model, a program can be seen as a flow of data which is processed by tasks, each task executing a distinct kernel. Failures within a program are nothing but a manifestation of a failed task, which can be identified as a local point of failure. This significantly simplifies the implementations of a resilient interface. The two intuitive choices of resiliency in AMT are task replay and task replicate. Task replay reschedules a failed task until it runs to completion or exhausts the number of replay trials. Task replicate schedules a single task multiple times (decided by the application developer) and, from the successfully completed tasks, determines the appropriate result.

The section \Romannum{2} talks about Related Work in AMT resilience. Section \Romannum{3} discusses about HPX and Resilience with HPX. Section \Romannum{4} discusses about the implementation details of implementing resilience. Section \Romannum{5} describes the benchmarks and Section \Romannum{6} discusses the results.

\section{Related Work}

Software based resilience for SPMD programs have been well studied and explored including but not limited to coordinated checkpoint and restart (C/R). Enabling resilience in AMT execution model has not been well studied despite the fact that the AMT paradigm facilitates an easier implementation. Subasi \textit{et al} ~\cite{b11, b12, b13} have discussed a combination of task replay and replicate with C/R for task-parallel runtime, OmpSs. For task repication, they suggested to defer launch of the third replica until duplicated tasks report a failure. This differs from our implementation, as we replicate the tasks and do not defer the launch of any task for future. For task replay, they depend on the errors triggered by the operating system. This approach, thus, assumes a reliable failure detection support by the operating system, which is not always available. We also found that automatic global checkpointing has been explored within the Kokkos ecosystem as well ~\cite{b14}.

Similar work has recently been explored in AMT with Habanero C ~\cite{b15}. The work, however, is based on on node resiliency. We plan to extend our work to provide distributed resiliency features. Furthermore, Our work implements the AMT resliency APIs with different characteristics. We also provide more control over the APIs by introducing multiple variants of a single resilience API. Finally, HPX is standards C++ conforming, giving application developers the least effort for the addition of resilience over their current non resilient code.

\section{Background}


\subsection{HPX}

HPX~\cite{heller2012,heller2013,kaiser2014,kaiser2015,heller2016} is a C++ standard library for 
distributed and parallel programming 
built on top of an asynchronous many-task 
(AMT) runtime system. Such AMT runtimes may provide a means for helping programming models to fully exploit 
available parallelism on complex emerging HPC architectures. The HPX programming model includes the 
following essential components: {\it (1)} an ISO C++ standard conforming API that enables wait-free asynchronous 
parallel programming, including futures, channels, and other asynchronization primitives; {\it (2)} an active global address 
space (AGAS) that supports load balancing via object migration; {\it (3)} an active-message networking layer that ships 
functions to the objects they operate on; {\it (4)} work-stealing lightweight task scheduler that enables finer-grained 
parallelization and synchronization. 

\subsection{Resilience in HPX}

In this work we assume that a ``failure'' is a manifestation of a failing task. 
A task is considered ``failing'' if it either throws an exception or if additional facilities 
(e.g. a user provided ``validation function'') identifies the computed result as being incorrect. 
This notion simplifies the implementation of resilience and makes HPX a suitable
platform to perform experiments  with resiliency APIs. We present two different 
ways to expose resiliency capabilities to the user: 

\textbf{Task Replay} is analogous to the Checkpoint/Restart mechanism found in 
conventional execution models. The key difference being localized 
fault detection. When the runtime detects an error it replays the failing 
task as opposed to completely rolling back of the entire program to the
previous checkpoint.

\begin{figure}[h]
\centering
  \includegraphics[width=0.8\linewidth]{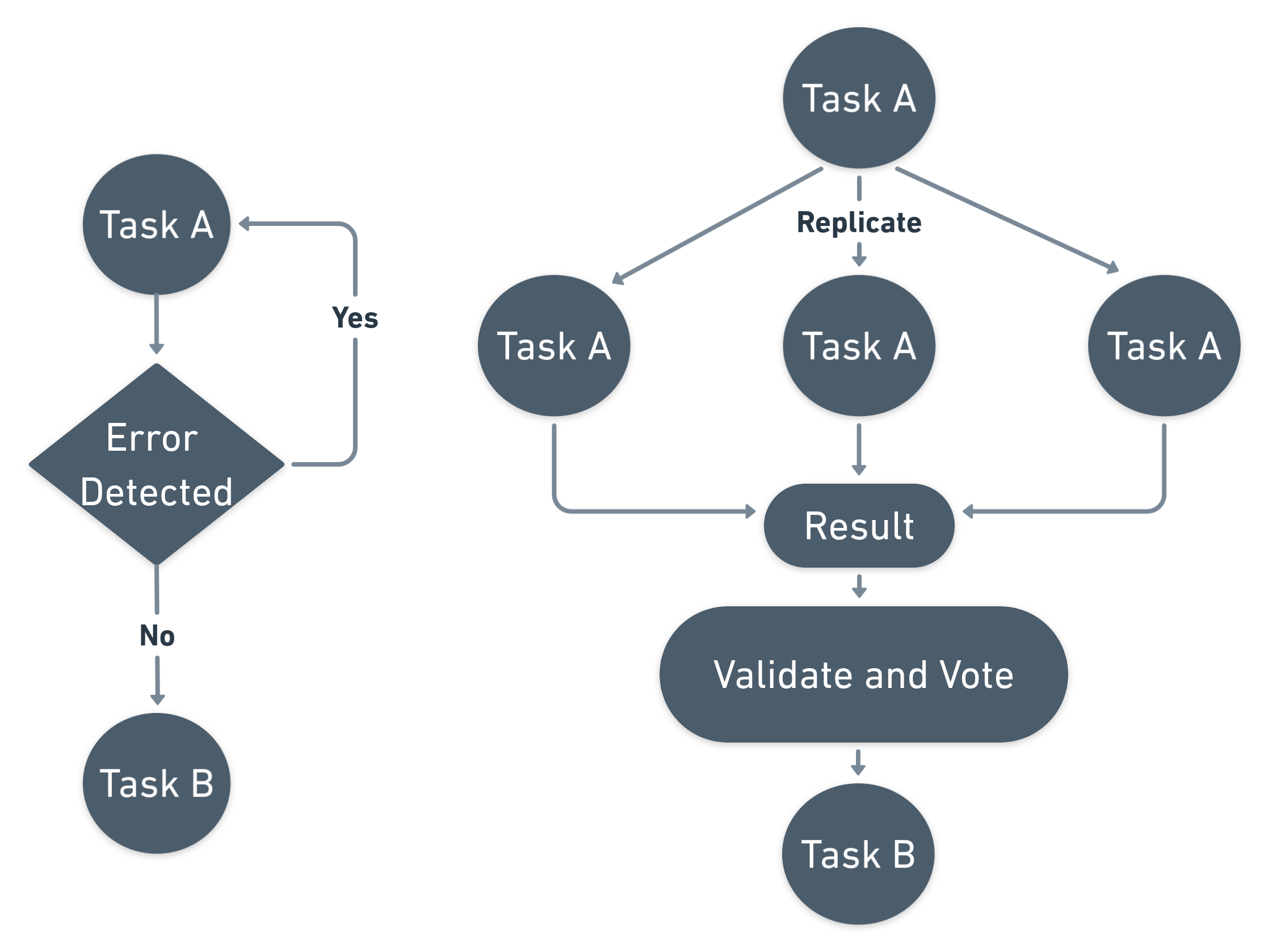}
  \caption{Task Replay and Task Replicate}
  \label{fig:replay}
\end{figure}

\textbf{Task Replicate}  is designed to provide reliability enhancements 
by replicating a set of tasks and evaluating their results to determine 
a consensus among them. This technique is most effective in 
situations where there are few tasks in the critical path of the DAG 
which leaves 
the system underutilized or where hardware or software 
failures may result in an incorrect result instead of an error. 
However, the drawback of this method is the additional computational 
cost incurred by repeating a task multiple times.

\section{Implementation Details}

The two main resiliency APIs explored are described below. All new functionalities
are implemented as extensions of the existing HPX {\it async} and {\it dataflow} API functions. This
enables a seamless migration of existing HPX codes to support the described resiliency features.

\subsection{Task Replay}\label{AA}

In this technique, a task is automatically replayed (re-run) up-to {\it N} times if an 
error is detected 
(see Listing~\ref{replay_api}).

(\romannum{1}) \textbf{Async and Dataflow Replay:} This version 
of task replay will catch user defined exceptions and automatically 
reschedule the task {\it N} times before re-throwing the
exception.

(\romannum{2}) \textbf{Async and Dataflow Replay Validate:} This 
version of replay adds a validation function to the API that is used to 
validate the individual results. It returns the first positively validated result.
If the validation fails or an exception is thrown, 
the task is replayed until no errors are encountered or the number 
of specified retries has been exceeded.


\begin{lstlisting}[frame=single, language=C++, caption={Task Replay API calls with variations. {\it N} represents the number of times the runtime system should attempt to reschedule the task, {\it F} is the function (task) to execute, {\it Args...} are the arguments to pass to {\it F}, {\it ValF} is the function to validate the results.}, label={replay_api}]
using namespace hpxr = hpx::resiliency;

hpxr::async_replay(N, F, Args...);
hpxr::dataflow_replay(N, F, Args...);

hpxr::async_replay_validate(N, ValF, F, Args...);
hpxr::dataflow_replay_validate(N, ValF, F, Args...);

\end{lstlisting}

\subsection{Task Replicate}\label{AA}

This feature launches {\it N} instances of a task concurrently. The 
function will take one of four code paths depending on the 
variation of the API (see Listing~\ref{replicate_api}):

(\romannum{1}) \textbf{Async and Dataflow Replicate:} 
This API returns the first  result that runs without errors.

(\romannum{2}) \textbf{Async and Dataflow Replicate Validate:} This
API additionally takes a function that validates the idnividual results. 
It returns the first result that is positively validated.

(\romannum{3}) \textbf{Async and Dataflow Replicate Vote:} This API 
adds a voting function to the basic replicate function. Many hardware or 
software failures are silent errors that do not interrupt the program flow.
The API determines the ``correct'' result by using the voting function allowing
to build a consensus. 

(\romannum{4}) \textbf{Async and Dataflow Replicate Vote Validate:} This 
combines the features of the previously discussed replicate APIs. Replicate 
vote validate allows a user to provide a validation function and
a voting function to filter results.
Any exceptions thrown during execution of the task are handled and are treated
as if the task failed. If all of the replicated tasks encounter 
an error, the last exception encountered while computing the task 
is re-thrown. If finite results are computed but fail the validation 
check, an exception is re-thrown.

\begin{lstlisting}[frame=single, language=C++, caption={Task Replicate API calls with variations. {\it N} is the number of replicate tasks to be launched concurrently, {\it F} is the function (task) to execute, {\it Args...} are the arguments to pass to {\it F}, {\it ValF} is the function to validate the results, and {\it VoteF} is the function to use to select the correct result.}, label={replicate_api}]
using namespace hpxr = hpx::resiliency;

hpxr::async_replicate(N, F, Args...);
hpxr::dataflow_replicate(N, F, Args...);

hpxr::async_replicate_validate(N, ValF, F, Args...);
hpxr::dataflow_replicate_validate(N, ValF, F, Args...);

hpxr::async_replicate_vote(N, VoteF, F, Args...);
hpxr::dataflow_replicate_vote(N, VoteF, F, Args...);

hpxr::async_replicate_vote_validate(N, VoteF, ValF, F, Args...);
hpxr::dataflow_replicate_vote_validate(N, VoteF, ValF, F, Args...);
\end{lstlisting}

\section{Benchmarks}

This section discusses the benchmark examples, the machine architecture and the HPX configuration.

\textbf{Machine:} All our benchmarks were run on a single node of NERSC's Supercomputer Cori. Each node has two sockets, each with Haswell Xeon E5-2698 v3 CPUs at 2.30GHz. While hyperthreading is enabled on the node, we have always run not more than one kernel thread per core (32 CPU threads). Each physical core has a dedicated L1 cache of 32KB and L2 cache of 256KB. Each socket has an L3 cache of 40MB shared between thirty two physical cores.

\textbf{HPX configuration:} We use Boost version 1.70.0, 
binutils version 2.32 and jemalloc version 5.2.0 for the
HPX builds. Boost and HPX were built with gcc 8.3 and all the 
benchmarks use gcc 8.3 as well~\cite{hpx_git}.

\textbf{Benchmarks:} We ran our experiments with two benchmarks. The
first benchmark was written to allow us to easily vary an artificial workload. 
The second benchmark application is a 1D stencil application that was 
adapted from a preexisting HPX example.

To ensure statistically relevant results, we ran all of the benchmarks 10 times and 
report the average timing as the benchmark's execution time. We did 
not include the initialization and shutdown costs in the measured execution time. 

\subsection{Artificial work loads}

This benchmark was written in order for the user to precisely 
control the task grain size and therefore correctly compute 
the overheads of the resiliency implementation. The benchmark 
calls a function 1,000,000 times and measures the execution time. 
The function takes several arguments including the task grain size 
and the error rate. The task grain size argument enables the user
to change the amount of ``work'' contained in each task.
In order to model random failures in the computation,
the function uses an error rate to adjust the percentage of 
tasks which will ``fail".

Within the function (see Listing~\ref{benchmark_task}), we added a timer to accurately measure 
the user specified task grain size and an atomic counter
to count the total number of failed tasks. Furthermore, 
the task may throw an exception to simulate ``failure'' based
on a probabilistic criterion.
In the case of the validate API, we 
compare the final computed result with our expected result. 
The benchmark measures the resilience capabilities 
of the system and the robustness of the APIs themselves.

\begin{lstlisting}[frame=single, language=C++, caption={Function body of a task run in the artificial benchmark}, label={benchmark_task}]
int universal_ans(uint64_t delay_ns, double error)
{
    std::exponential_distribution<> dist(error);
    double num = dist(gen);
    bool error_flag = false;
    if (num > 1.0)
    {
        error_flag = true;
        ++counter;
    }
    uint64_t start = high_resolution_clock::now();
    while (true)
    {
        uint64_t now = high_resolution_clock::now();
        if (now - start >= delay_ns)
        {
            if (error_flag) throw std::exception();
            break;
        }
    }
    return 42;
}
\end{lstlisting}

\subsection{1D Stencil}

For this benchmark, we ported the 1D stencil code from HPX to enable resilience while adding multiple time advancing steps per iteration on a stencil. This benchmark accurately measures the overheads that one can observe while using dataflow resilient variants. The overheads encountered with dataflow arise from two factors. 
First, the overheads introduced from the creation of 
the dataflow object, and second the sequence in 
which the futures passed to the dataflow become ready.
A dataflow waits for all provided futures to become ready, 
and then executes the specified function. 

This benchmark solves a linear advection equation. 
The task decomposition, Lax-Wendroff stencil, and checksum operations are as described in previous work~\cite{b15}.
Each task has three dependencies, the subdomain the current task works 
on and the left and right neighboring subdomains.
The stencil is advanced through multiple time steps in each task by reading an extended ``ghost region'' of data values from each neighbor, which helps reducing overheads and latency effects.

We run the benchmark with two cases that we call 1D stencil case A and 1D stencil case B. The case A uses 128 subdomains
each with 16,000 data points, it runs for 8,192 iterations with 128 time steps per iteration. The case B uses 256
subdomains each with 8,000 data points, it runs for 8,192 iterations with 128 time steps per iteration. The performance
depends upon the task grain size, i.e. the subdomain size, the number of time steps advanced per iteration, and the number
of dataflow operations involved, i.e. number of subdomains and iterations. The performance overhead of resilience 
depends upon the
number of tasks. Case A invokes a total of 1,048,576 tasks while Case B involves 2,097,152 tasks. 

\subsection{Injecting errors}

Errors injected within the applications are artificial 
and not a reflection of any computational or memory 
errors. We use an exponential distribution 
function to generate an exponential curve signature such that 
the probability of errors is equal to $e^{-x}$, where x is the 
error rate factor. For example, an error rate of 1 will have the 
probability of introducing an error within a task equal to $e^{-1}$ or 0.36.

\section{Results}

This section discusses the empirical results for the benchmarks described in Section \Romannum{5}.

\subsection{Async Replay and Replicate}

Resilient variants of {\it async} were measured with the artificial benchmarks 
to compute the overheads. The overheads introduced by our implementation 
are listed in Table~\ref{overheads}. The observed implementation overheads are 
very small and are often comparable to the measurement accuracy. 

\begin{table}[h!]
  \centering
    \begin{tabular}{|c|c|c|c|c|c|c|}
    \hline
    \begin{tabular}[c]{@{}c@{}}No. of \\ Cores\end{tabular} & \multicolumn{2}{c|}{\begin{tabular}[c]{@{}c@{}}Async Replay and\\ Replay Validate \\ (in $\mu$s)\end{tabular}} & \multicolumn{4}{c|}{\begin{tabular}[c]{@{}c@{}}Async Replicate, Replicate Validate,\\ Replicate Vote and Replicate Vote\\ Validate (in $\mu$s)\end{tabular}} \\ \hline
    1                                                       & 0.792                                                & 0.774                                               & 0.985                                & 0.986                                & 0.987                                & 1.023                               \\ \hline
    4                                                       & 0.251                                                & 0.263                                               & 0.161                                & 0.165                                & 0.161                                & 0.163                               \\ \hline
    8                                                       & 0.145                                                & 0.150                                               & 0.078                                & 0.078                                & 0.078                                & 0.082                               \\ \hline
    16                                                      & 0.080                                                & 0.085                                               & 0.034                                & 0.034                                & 0.034                                & 0.036                               \\ \hline
    32                                                      & 0.057                                                & 0.058                                               & 0.017                                & 0.017                                & 0.017                                & 0.016                               \\ \hline
    \end{tabular}
    \caption{Amortized overheads per task of resilient {\it async} variants with a task grain size of 200$\mu$s \label{overheads}}
  \end{table}  

\begin{figure}[h!]
  \centering
  \begin{subfigure}[b]{1.0\linewidth}
    \includegraphics[width=\linewidth]{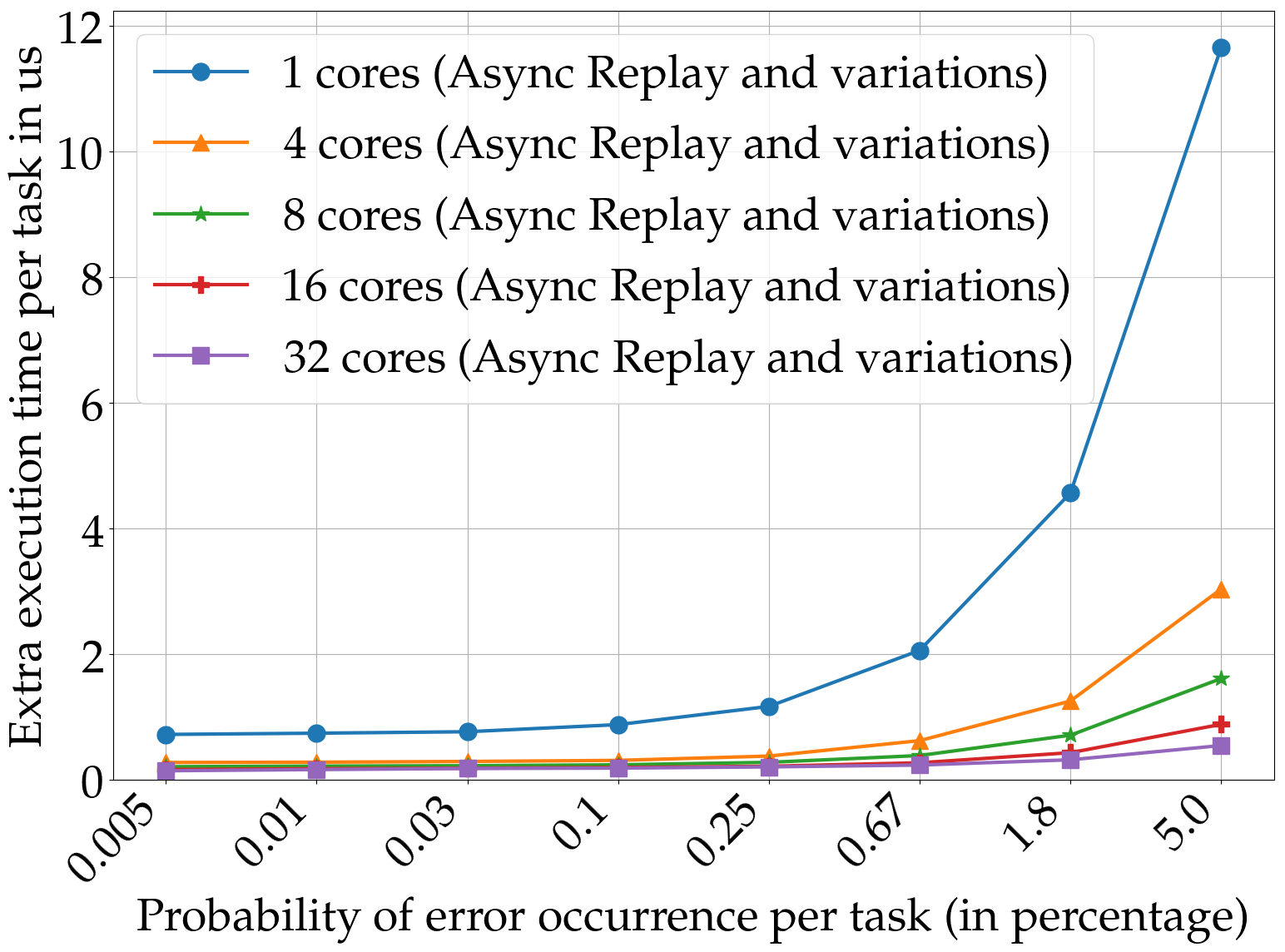}
    \caption{Async Replay: Extra execution time per task vs. Probability of error occurrence. \label{replay_errors}\newline}
  \end{subfigure}
  \begin{subfigure}[b]{1.0\linewidth}
    \includegraphics[width=\linewidth]{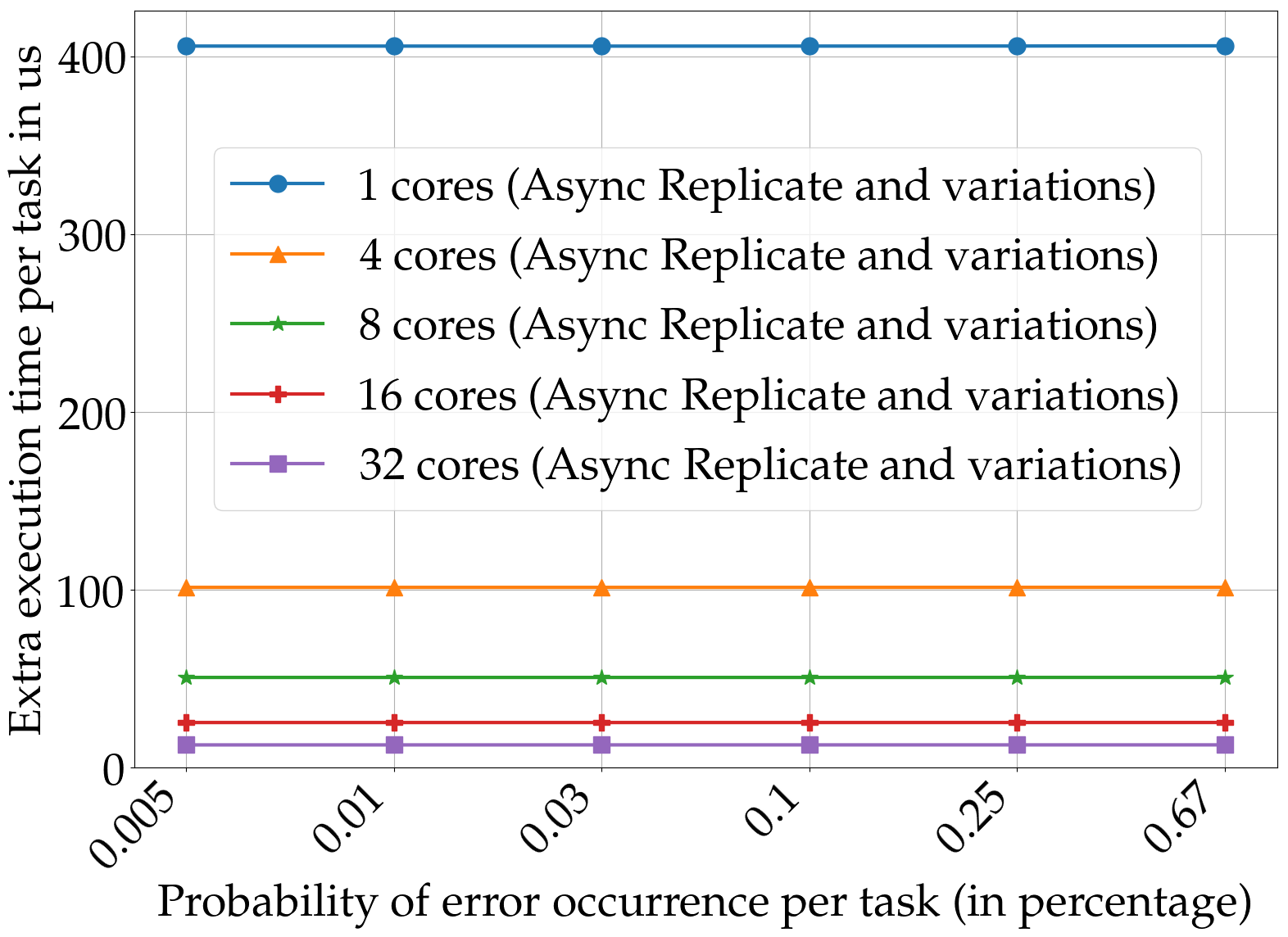}
    \caption{Async Replicate: Extra execution time per task vs. Probability of error occurrence \label{replicate_errors}}
  \end{subfigure}
  \caption{Extra execution time per task for task grain size of 200$\mu$s.}
  \label{fig:async_extra}
\end{figure}

Replicate variants of async behave similarly. The overheads
for replicate depends upon the number of replications. 
Other minor overheads such as 
vector memory allocation and the number of push back function 
calls are also dependent on the number of replicates requested. 
With sufficient compute resources available, 
one can expect these overheads to be within the 
margin of error. For a task of 200$\mu$s making three replicates, 
there is thrice the number of tasks involved compared to its replay 
counterpart (see also Figure~\ref{replicate_errors}). Because these extra tasks are independent of each 
other, the overheads are quickly amortized when there are 
available resources than its replay counterpart, which will 
not be able to take advantage of the added parallelism as 
efficiently. The minor differences in overheads 
between resilient variants arises from the underlying 
implementation, some requiring more boilerplate code 
to be executed than others.


The trend is similar when the applications are 
injected with artificial errors. When errors are encountered, 
the resilient logic is activated and behaves as specified. For cases 
with low probability of failures, we see that amortized overheads 
of async replay and variants are still small enough to be hidden 
by system noise (see also Figure~\ref{replay_errors}). 
For instance, the overheads, in the worst case scenario for an error 
rate of 5\%, is about 
0.54$\mu$s or 0.27\% of the total overhead per task.
Given that the probability of failure within a machine will not be
more than a percent in most cases, it is safe to assume that 
async replay introduces no measurable overheads for applications utilizing
the feature. 
Taken together, the presented results indicate that 
these resilient features 
will not incur any meaningful execution time costs.

For cases that use async replicate and variants, we observe 
that the amortized overhead per task stand at 6.4\% for task 
size of 200$\mu$s, which is significantly larger. This is expected as 
the overheads include running more duplicate tasks. 
A minor implementation overhead is also present, though 
dwarfed by the time it takes to repeatedly run a task itself.
The graph is a straight line, as expected, since every task 
is replicated three times irrespective of any encountered errors (see Figure~\ref{replicate_errors}). 
With these costly overheads, this resiliency feature is only 
recommended in portions of code which are starved for work (i.e. there are sufficient 
computational resources available)
or for critical portions of code, where computed results have to be guaranteed to be correct.

\subsection{1D stencil}

1D stencil example was designed to check the {\it dataflow} resiliency capabilities. The aim of this benchmark was to observe the extra execution time for different stencil parameters. Table~\ref{1d_stencil_times} summarizes the measured results for the case of no failures, comparing the new API variations with a base-line version that is based on basic HPX {\it dataflow} facilities. 
\begin{table}[h!]
\centering
  \begin{tabular}{|c|c|c|c|c|}
  \hline
  \begin{tabular}[c]{@{}c@{}}1D \\ stencil\end{tabular} & \begin{tabular}[c]{@{}c@{}}Pure \\ Dataflow\\ (in s)\end{tabular} & \begin{tabular}[c]{@{}c@{}}Replay \\ without\\ checksums\\ (in s)\end{tabular} & \begin{tabular}[c]{@{}c@{}}Replay \\ with\\ checksums\\ (in s)\end{tabular} & \begin{tabular}[c]{@{}c@{}}Replicate \\ without\\ checksums\\ (in s)\end{tabular} \\ \hline
  Case A                                                & 46.564                                                            & 47.315                                                                         & 46.869                                                                      & 135.871                                                                           \\ \hline
  Case B                                                & 47.267                                                            & 49.756                                                                         & 49.268                                                                      & 139.242                                                                           \\ \hline
  \end{tabular}
  \caption{1D stencil: Execution time in case of no failures for case A and case B, where case A utilizes 128 subdomains each with 16000 data points and case B utilizes 256 subdomains each with 8000 data points, each case iterating 8192 times with 128 times steps per iteration. \label{1d_stencil_times}}
\end{table}


Implementing resiliency using the {\it dataflow} variations with 1D stencil case A introduces 1.5\% and 0.4\% overheads
for replay without and with checksums respectively. Similarly, with case B we observe 5\% and 4.1\% overheads for replay
without and with checksums respectively. These numbers are certainly higher than its {\it async} variants. This is
expected given that dataflow waits for all futures to become ready before executing the function. We currently cannot explain 
why the benchmarks with checksums runs (slightly) faster. The time difference in execution might arise due to lower
overheads of synchronization within dataflow. 
Resiliency with replication are low as expected as well, which can be attributed owing to
efficient parallelism and cache effects.

When we inject 
errors within the code, we see similar trends as those we observed 
in the case of its async counterpart. For low failure rates, we observe 
that the overheads are about the same as the implementation overheads. 
As expected we see a spike in overheads as the probability of failure 
increases with 5.9\% and 6.9\% overheads for Case A and 8.5\% and 
9.6\% for Case B.

\begin{figure}[h!]
  \centering
  \begin{subfigure}[b]{1.0\linewidth}
    \includegraphics[width=\linewidth]{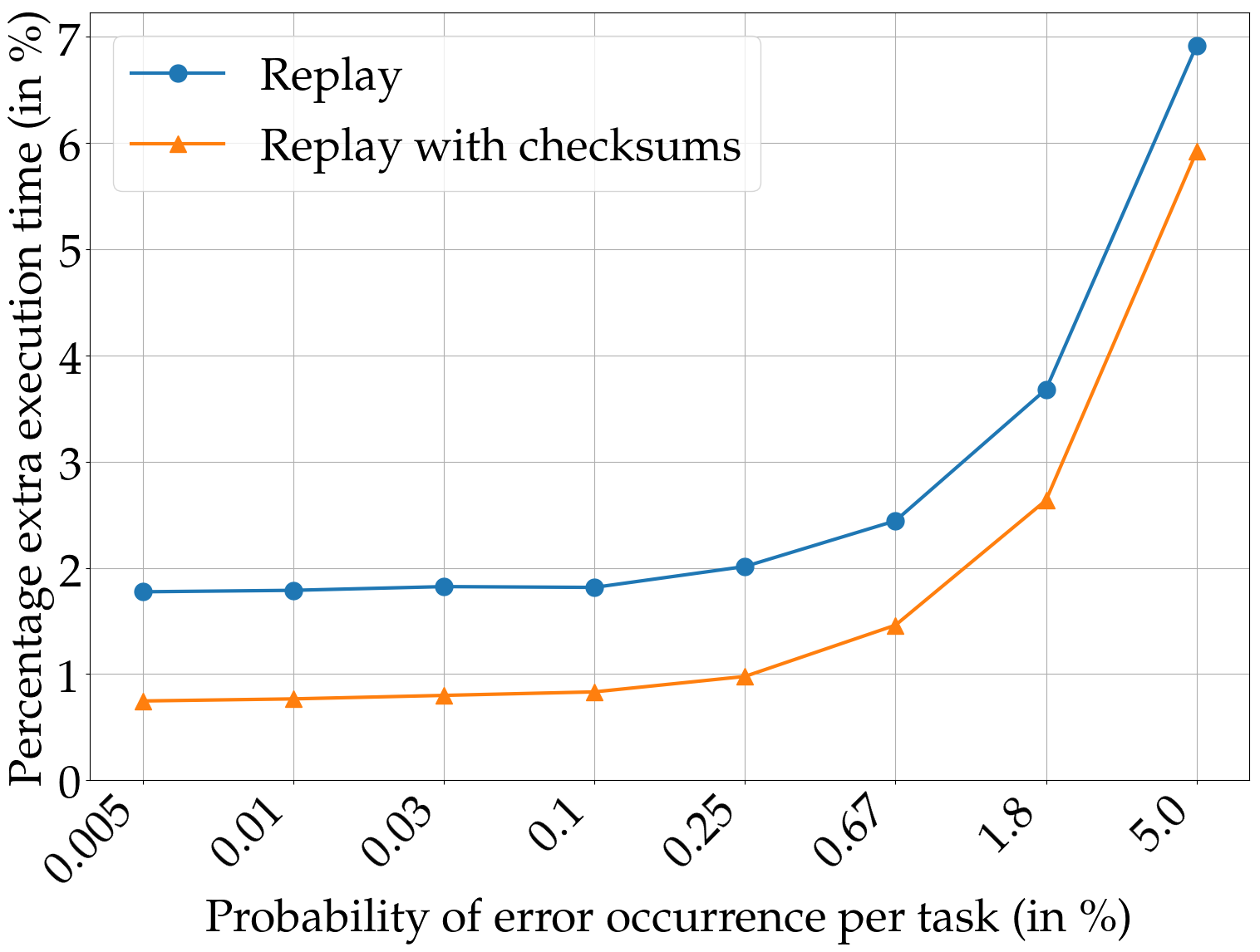}
    \caption{1D stencil case A: Percentage extra execution time vs. Probability of error occurrence. \label{replay_dataflow_errors}\newline}
  \end{subfigure}
  \begin{subfigure}[b]{1.0\linewidth}
    \includegraphics[width=\linewidth]{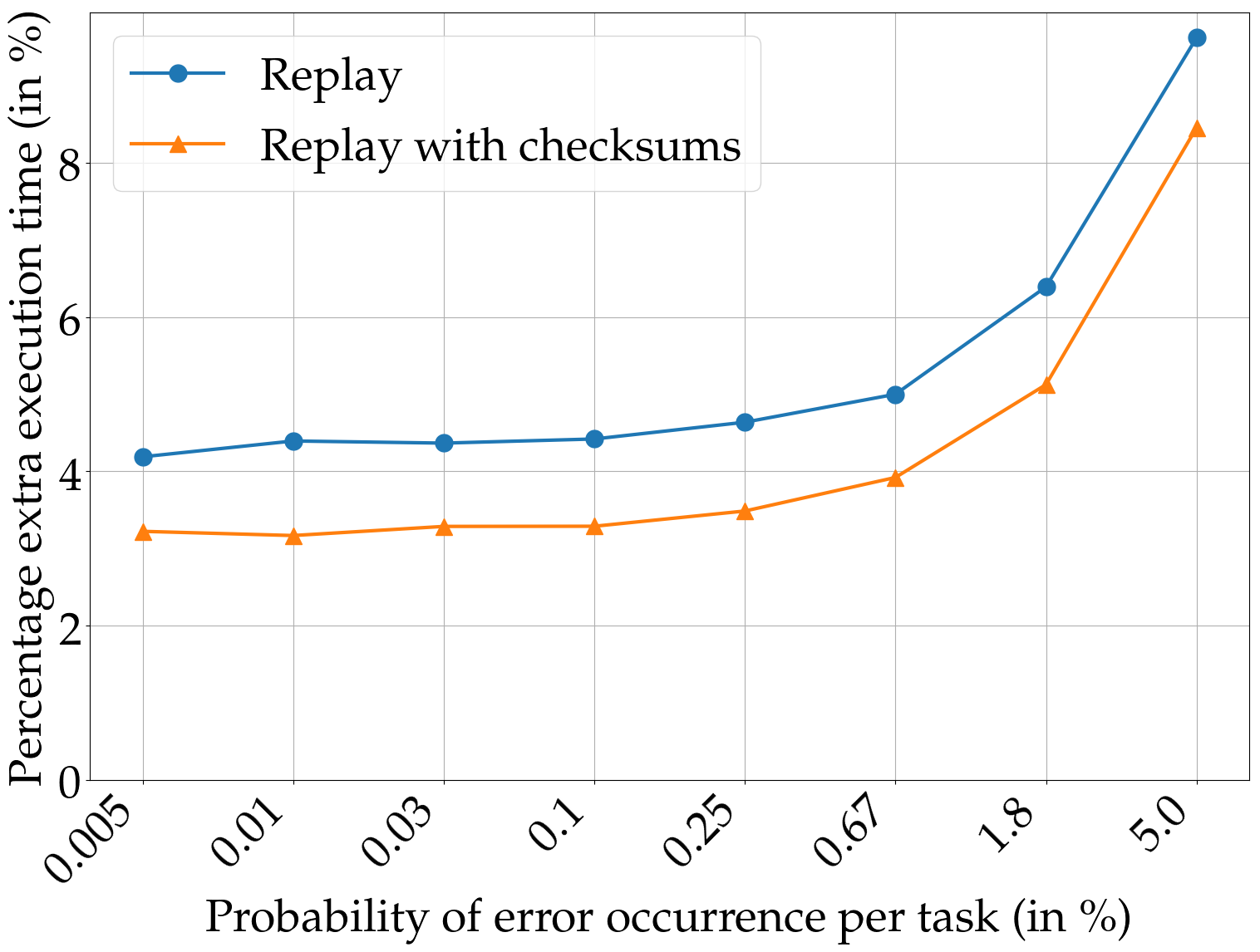}
    \caption{1D stencil case B: Percentage extra execution time vs. Probability of error occurrence. \label{replicate_dataflow_errors}}
  \end{subfigure}
  \caption{1D stencil case A and case B: case A  utilizes  128  subdomains  each  with  16000  doubles and case B utilizes 256 subdomains each with 8000 doubles, and  each  case  iterates  8192  times  with  128  times  steps  periteration.}
  \label{fig:stencil_extra}
\end{figure}

\section*{Conclusion}
In this paper, we implemented two resiliency APIs in HPX: task replay and task replication. 
Task replay reschedules a task up to n-times until a valid output is returned. 
Task replication runs a task n-times concurrently. 
We demonstrate that only minimal overheads are incurred when utilizing these resiliency features for 
work loads where the task size is greater than 200 $\mu$s. We also show that most of the 
added execution time arises from the replay or replication of the tasks themselves and not from the
implementation of the APIs.


Furthermore, as the new APIs are designed as extensions to the existing HPX {\it async} 
facilities that are fully conforming to the C++ standard, these features will be easy enough to 
embrace and enable a seamless migration of existing code. 
Porting a non resilient 
application to its resilient counterpart will require minimal changes, 
along with the implementation of validation/vote functions, wherever necessary. 
This removes the necessity of costly code re-writes as well 
as time spent learning new APIs.

Finally, we developed multiple resilient APIs that are designed
to cater to the specific needs of 
an application. This allows developers the option to choose 
the optimal resilient function with the least disruption 
to efficiency.

\section*{Future Work}

The current implementation of resiliency is limited to intra node parallelism.
We plan to extend the presented resiliency facitities to the distributed case while
maintaining the straightforward API. We expect that both -- task replay and task 
replicate -- can be seamlessly extended to the distributed use case by introducing special
executors that will manage the aspects of resiliency and task distribution across nodes.


The current implementation enables to have both task replay 
and task replicate within the code independently. Task replicate can be made 
more robust by adding task replay within its implementation 
allowing any failed replicated task to replay until its computed 
without error detection. This will allow for finer consensus in 
case of soft failures within the system.

\section*{Acknowledgment}

Sandia National Laboratories is a multimission laboratory managed and operated by National Technology \& Engineering Solutions of Sandia, LLC, a wholly owned subsidiary of Honeywell International Inc., for the U.S. Department of Energy's National Nuclear Security Administration (NNSA) under contract DE-NA0003525. This work was funded by NNSA's Advanced Simulation and Computing (ASC) Program. This paper describes objective technical results and analysis. Any subjective views or opinions that might be expressed in the paper do not necessarily represent the views of the U.S. Department of Energy or the United States Government.

\input{appendix.reproducibility.tex}

\end{document}

%% file: appendix.reproducibility.tex
\section{Artifact Description: Implementing Software Resiliency in HPX forExtreme Scale Computing}


\subsection{Abstract}

In this section we describe the configuration and environment needed to run our experiments. The experiments are run on a single node of NERSC Cori supercomputer. We present the steps to build the software on Cori and repeat our experiments and how to gather the performance results.

\subsection{Description}

\subsubsection{Check-list (artifact meta information)}


{\small
\begin{itemize}
  \item {\bf Program: HPX}
  \item {\bf Compilation: GCC 8..3}
  \item {\bf Hardware: Cori}
  \item {\bf Run-time state: }
  \item {\bf Output: Benchmark results in text}
  \item {\bf Publicly available?: Yes}
\end{itemize}
}

\subsubsection{How software can be obtained (if available)}
HPX can be obtained from \url{https://github.com/STEllAR-GROUP/hpx}. Resiliency features are available as an HPX module within HPX itself.


\subsubsection{Hardware dependencies}
Access to Cori can be requested through NERSC.

\subsubsection{Software dependencies}
\label{software-dependencies}
The following modules need to be loaded to build HPX:
\begin{itemize}
\item{gcc/8.3.0}
\item{boost/1.70.0}
\item{cmake/3.10.2}
\end{itemize}

The following software need to be built from their source since they are not available on Cori:
\begin{itemize}
\item{jemalloc. The latest version of CMake can be downloaded from \url{https://github.com/jemalloc/jemalloc}.}
\end{itemize}

\subsection{Installation}
\begin{itemize}
\item{HPX

\begin{lstlisting}
cmake
  -H<PATH_TO_HPX_CODE>
  -B<PATH_TO_HPX_BUILD>
  -DCMAKE_BUILD_TYPE=Release
  -DCMAKE_C_COMPILER=gcc
  -DCMAKE_CXX_COMPILER=g++
  -DBOOST_ROOT=<PATH_TO_BOOST>
  -DBoost_NO_SYSTEM_PATHS=True
  -DHPX_WITH_EXAMPLES=False
cmake --build <PATH_TO_HPX_BUILD>
\end{lstlisting}
}
\end{itemize}


\subsection{Running Benchmarks}
\begin{itemize}
\item{Get a job from Cori for a single node haswell and ensure all required modules listed in section \ref{software-dependencies} are loaded.}
\item{Copy the script files provided at \url{https://github.com/STEllAR-GROUP/hpxr_data/tree/master/benchmark_scripts} to the binary directory. Execute the scripts to generate the results.}
\item{Repeat the same for all the shell scripts}
\end{itemize}


\subsection{Evaluation and expected result}

Scripts to generate graphs on the returned results are provided at \url{https://github.com/STEllAR-GROUP/hpxr_data/tree/master/benchmark_results/graphs}. Copy them to the binary directory and run the python files. To successfully generate the graphs make sure that the following are available:
\begin{itemize}
    \item {python 2.7}
    \item {matplotlib}
    \item {Tkinter}
\end{itemize}

The emperical results that we achieved running the scripts described above are provided at \url{https://github.com/STEllAR-GROUP/hpxr_data/tree/master/benchmark_results/benchmarks}. Running the python scripts on them will result in the graphs used in the paper.


%% file: paper.bbl
\begin{thebibliography}{00}
\bibitem{b1}  P. Kogge et al., ExaScale Computing Study: Technology Challenges in Achieving Exascale Systems, tech. report TR-2008-13, Dept. of Computer Science and Eng., Univ. of Notre Dame, 2008
\bibitem{b2} J. T. Daly, “A higher order estimate of the optimum checkpoint interval
for restart dumps,” Future Gener. Comput. Syst., vol. 22, no. 3, pp.
303–312, Feb. 2006.
\bibitem{b3} J. Duell, P. Hargrove, and E. Roman, “The design and implementation
of berkeley lab’s linux checkpoint/restart,” Lawrence Berkeley National
Laboratory, Tech. Rep. LBNL-54941, 2002.
\bibitem{b4} K. Li, J. Naughton, and J. Plank, “Low-latency, concurrent checkpointing for parallel programs,” Parallel and Distributed Systems, IEEE Transactions on, vol. 5, no. 8, pp. 874–879, Aug 1994.
\bibitem{b5} A. Moody, G. Bronevetsky, K. Mohror, and B. de Supinski, “Design, modeling, and evaluation of a scalable multi-level checkpointing system,” in High Performance Computing, Networking, Storage and Analysis (SC), 2010 International Conference for, 2010, pp. 1–11.
\bibitem{b6} J. S. Plank, K. Li, and M. A. Puening, “Diskless checkpointing,” IEEE Transactions on Parallel and Distributed Systems, vol. 9, no. 10, pp. 972–986, October 1998.
\bibitem{b7} E. Roman, “Survey of checkpoint/restart implementations,” Lawrence Berkeley National Laboratory, Tech. Rep. LBNL-54942, 2002.
\bibitem{b8} M. Gamell, D. S. Katz, H. Kolla, J. Chen, S. Klasky, and M. Parashar, “Exploring automatic, online failure recovery for scientific applications at extreme scales,” in Proceedings of the International Conference on High Performance Computing, Networking, Storage and Analysis, ser. SC ’14, 2014.
\bibitem{b9} K. Teranishi and M. A. Heroux, “Toward local failure local recovery resilience model using MPI-ULFM,” in Proceedings of EuroMPI/Asia’14, the 21st European MPI User Group Meeting, 2014.
\bibitem{b10} A. Guermouche, T. Ropars, E. Brunet, M. Snir, and F. Cappello, “Uncoordinated checkpointing without domino effect for send-deterministic mpi applications,” in Parallel Distributed Processing Symposium (IPDPS), 2011 IEEE International, 2011, pp. 989–1000.
\bibitem{b11} O. Subasi et al., “Nanocheckpoints: A task-based asynchronous datafow framework for efficient and scalable checkpoint/restart,” in 2015 23rd Euromicro International Conference on Parallel, Distributed, and Network-Based Processing, pp. 99–102, March 2015.
\bibitem{b12} O. Subasi et al., “A runtime heuristic to selectively replicate tasks for application-specific reliability targets,” in 2016 IEEE International Conference on Cluster Computing (CLUSTER), pp. 498–505, Sept 2016.
\bibitem{b13} O. Subasi et al., “Designing and modelling selective replication for fault-tolerant hpc applications,” in 2017 17th IEEE/ACM International Symposium on Cluster, Cloud and Grid Computing (CCGRID), pp. 452–457, May 2017.
\bibitem{b14} H. Carter Edwards et.al., “Kokkos: Enabling Manycore Performance Portability through Polymorphic 
Memory Access Patterns'', Journal of Parallel and Distributed Computing, 2014, vol. 74, num. 12, pp. 3202-3216.

\bibitem{b15} K. Teranishi et al., “ASC CSSE level 2 milestone \#6362: Resilient asynchronous many-task programming model,” Sandia National Laboratories, Tech. Rep. SAND2018-9672, 2018.

\bibitem{heller2012} Thomas Heller et.al., “Application of the ParalleX Execution Model to Stencil-Based Problems'', in Computer Science - Research and Development, 2012, vol. 28, num. 2-3, pp. 253-261.
\bibitem{heller2013} Thomas Heller et.al., “Using HPX and LibGeoDecomp for Scaling 
HPC Applications on Heterogeneous Supercomputers'', ScalA 2013, SC Workshop.
\bibitem{kaiser2015} Hartmut Kaiser et.al., “Higher-level Parallelization for Local and 
Distributed Asynchronous Task-based Programming'', ESPM 2015, SC Workshop.
\bibitem{heller2016} Thomas Heller et.al., “Closing the Performance Gap with Modern C++'', E-MuCoCoS 2016, ISC Workshop.

\bibitem{kaiser2014} H.Kaiser et.al., {“HPX: A Task Based Programming Model in a Global Address Space}'',
International Conference on Partitioned Global Address Space Programming Models (PGAS), 2014, art. id 6.

\bibitem{hpx_git} HPX GitHub repository, {STE$||$AR Group}, \url{https://github.com/STEllAR-GROUP/hpx}, 2019, Available under the Boost Software License 1.0.

\newpage
\end{thebibliography}
